\documentclass[%
 aip,
 jcp,%
 amsmath,amssymb,
 reprint,%
]{revtex4-1}

\usepackage{graphicx}
\usepackage{dcolumn}
\usepackage{bm}
\usepackage{color} 
\usepackage{amsmath,amssymb}
\usepackage{epsfig}
\usepackage{bbold}
\usepackage{url}
\usepackage{subfig}
\usepackage{lineno}
\usepackage{ulem}
\usepackage{amsmath,amssymb}
\usepackage[colorlinks=true,citecolor={blue},urlcolor={red},linkcolor={blue}]{hyperref}

\begin{document}
\title{Time-dependent optimized coupled-cluster
method for multielectron dynamics IV: Approximate consideration of the triple excitation amplitudes}
\author{Himadri Pathak}%
\email{pathak@atto.t.u-tokyo.ac.jp}
\affiliation{Department of Nuclear Engineering and Management, School of Engineering, The University of Tokyo, 7-3-1 Hongo, Bunkyo-ku, Tokyo 113-8656, Japan}%
\author{Takeshi Sato}
\email{sato@atto.t.u-tokyo.ac.jp}
\affiliation{Department of Nuclear Engineering and Management, School of Engineering, The University of Tokyo, 7-3-1 Hongo, Bunkyo-ku, Tokyo 113-8656, Japan}
\affiliation{Photon Science Center, School of Engineering, The University of Tokyo, 7-3-1 Hongo, Bunkyo-ku, Tokyo 113-8656, Japan}
\affiliation{Research Institute for Photon Science and Laser Technology, The University of Tokyo, 7-3-1 Hongo, Bunkyo-ku, Tokyo 113-0033, Japan}
\author{Kenichi L. Ishikawa}
\email{ishiken@n.t.u-tokyo.ac.jp}
\affiliation{Department of Nuclear Engineering and Management, School of Engineering, The University of Tokyo, 7-3-1 Hongo, Bunkyo-ku, Tokyo 113-8656, Japan}
\affiliation{Photon Science Center, School of Engineering, The University of Tokyo, 7-3-1 Hongo, Bunkyo-ku, Tokyo 113-8656, Japan}
\affiliation{Research Institute for Photon Science and Laser Technology, The University of Tokyo, 7-3-1 Hongo, Bunkyo-ku, Tokyo 113-0033, Japan}
\begin{abstract}
We present a cost-effective treatment of the triple excitation amplitudes in the
time-dependent optimized coupled-cluster (TD-OCC) framework called  TD-OCCDT(4)
for studying intense laser-driven multielectron dynamics.
It considers triple excitation amplitudes
correct up to fourth-order in many-body perturbation theory and achieves a computational scaling of $O(N^7)$, with $N$ being the number of active orbital functions. 
This method is applied to the electron dynamics in Ne and Ar atoms exposed to an intense near-infrared laser pulse with various intensities.
We benchmark our results against the time-dependent complete-active-space self-consistent field (TD-CASSCF), time-dependent optimized coupled-cluster with double and triple excitations (TD-OCCDT),
time-dependent optimized coupled-cluster with double excitations (TD-OCCD),
and the time-dependent Hartree-Fock (TDHF) methods to understand how this approximate scheme performs 
in describing nonperturbatively nonlinear phenomena, such as 
field-induced ionization and high-harmonic generation.
We find that the TD-OCCDT(4) method performs equally well as the TD-OCCDT method, almost perfectly reproducing the results of fully-correlated TD-CASSCF with a more favorable computational scaling.
\end{abstract}
\date{\today}
\maketitle
\section{Introduction}
The advance in experimental techniques for high-intensity ultrashort optical
pulses has contributed substantially to enrich strong-field physics and attosecond science \cite{itatani2004, corkum2007attosecond, krausz2009attosecond, baker2006probing}.
It is also showing promise to measure and manipulate the motions of electrons in a quantum many-body system \cite{goulielmakis2010real, sansone2010electron}.
One of the key processes in attosecond physics is high-order harmonic generation (HHG) \cite{antoine1996attosecond, calegari2014}.
It is a highly nonlinear frequency conversion process that delivers ultrashort coherent light pulses
in the extreme-ultraviolet (XUV) to the soft x-ray regions interacting with laser pulses of intensity $\gtrsim 10^{14}\,{\rm W/cm}^2$ 
in the visible to the mid-infrared spectral range.
The HHG spectra consist of a plateau where the intensity of the emitted radiation remains nearly constant up to many orders,
followed by a sharp cutoff, beyond which harmonics are ceased to observe \cite{kli2010}.

The time-dependent Schr{\"o}dinger equation (TDSE) provides the most rigorous theoretical 
description of all these dynamical phenomena \cite{parker1998intense, parker2000time, ishikawa2005above, feist2009probing, ishikawa2012competition, sukiasyan2012attosecond, 
vanroose2006double, horner2008classical}.
However, it remains a major challenge to achieve the exact numerical solution of TDSE for more than two-electron systems due to its high dimensionality.
Thus, single-active electron (SAE) approximations have been widely used \cite{krause1992jl, kulander1987time}, in which only the outermost electron is explicitly treated.
Though useful in obtaining a qualitative insight into different strong-field phenomena, SAE is intrinsically unable to treat the electron correlation in the field-induced multielectron dynamics.

Therefore, the development of various time-dependent, electron-correlation methods and the refinement of the existing ones
is ongoing for a better understanding of intense-laser-driven multielectron dynamics.
See Ref.~\citenum{ishikawa2015review} for a review on various wavefunction-based methods for the study of laser-driven electron dynamics,
and Ref.~\citenum{Lode:2020} for multiconfiguration approaches for indistinguishable particles. 
The multi-configuration time-dependent Hartree-Fock (MCTDHF) method \cite{caillat2005correlated, kato2004time, nest2005multiconfiguration,
haxton2011multiconfiguration, hochstuhl2011two}
and time-dependent
complete-active-space self-consistent-field (TD-CASSCF) method \cite{sato2013time, sato2016time} are the best known for the purpose.
Both these methods are based on the full configuration interaction (FCI) expansion of the wavefunction,
$\Psi(t)=\sum_{\bm I}C_I(t)\Phi_{\bm I}(t)$, where both CI coefficients $\{C_{\bm I}(t)\}$ and {\it occupied}
spin-orbitals $\{\psi_p(t)\}$ constituting Slater determinants $\{\Phi_{\bm I}(t)\}$ are time-dependent and propagated in time according to the 
time-dependent variational principle (TDVP).
The TD-CASSCF method broadens the applicability by flexibly classifying occupied orbital space into frozen-core (do not participate in dynamics),
dynamical-core (do participate in the dynamics remaining doubly occupied), and active (fully correlated description of the active electrons) subspaces.
By such flexible classification, it helps to get rid of a large number of chemically or physically inert electrons from the correlation calculation and thereby enhance the applicability.
However, these methods scale factorially with the number of correlated electrons, restricting large scale applications.
By limiting the configuration interaction (CI) expansion of the wavefunction up to an affordable level,
various less demanding methods \cite{miyagi2013time, miyagi2014time, haxton2015two, sato2015time, erik2018}
like the time-dependent restricted-active-space self-consistent field (TD-RASSCF) \cite{miyagi2013time, miyagi2014time}
and time-dependent occupation-restricted multiple-active-space (TD-ORMAS) \cite{sato2015time} have been developed.
These methods have proven to be of great utility \cite{imam} even though not  size-extensive.

On the other hand, the coupled-cluster method is the most celebrated many-body method while dealing with the electron correlation.
It is size-extensive at any given level of truncation, scales polynomially, and has proven its effectiveness
on a wide range of problems in the stationary theory \cite{helgaker2014molecular, kummel:2003, shavitt:2009}.
Therefore, we have developed its time-dependent extension using optimized orthonormal orbitals,
called the time-dependent optimized coupled-cluster (TD-OCC) method \cite{sato2018communication}.
The TD-OCC method is the time-dependent formulation of the stationary orbital optimized coupled-cluster method \cite{scuseria1987optimization, sherrill1998energies, krylov1998size},
and we have considered double and triple excitation amplitudes (TD-OCCDT) in our implementation.
Here again, both orbitals and amplitudes are time-dependent and propagated in time according to TDVP. 
We applied our method to simulate HHG spectra and strong-field ionization in Ar. 
The TD-OCCDT produced virtually identical results with the 
TD-CASSCF method in a polynomial cost-scaling
within the same chosen active orbital subspace \cite{sato2018communication}.
In an earlier development, Kvaal reported an orbital adaptive
time-dependent coupled-cluster (OATDCC) method \cite{kvaal2012ab} built upon the work of Arponen using biorthogonal orbitals \cite{arponen1983variational}.
However, this work does not address applications to laser-driven dynamics.
We also take note of the very initial attempts of the time-dependent coupled cluster \cite{schonhammer1978schonhammer, hoodbhoy1978time, hoodbhoy1979time}, and its applications using time-independent orbitals \cite{GUHA199184, prasad1988time, sebastian1985correlation, huber2011explicitly, 
pigg2012time, nascimento2016linear}.
See Ref.~\citenum{kristiansen2020numerical} for the discussion on numerical stability of time-dependent coupled-cluster methods with time-independent orbitals, and Refs.\citenum{pedersen2019symplectic,  pedersen2020interpretation} for symplectic integrator and interpretation of the time-dependent coupled-cluster method.
The TD-OCCDT method scales as $O(N^8)$, with $N$ being the number of active orbitals.

To further reduce the computational scaling and thereby enhance the applicability
to larger chemical systems, we have developed
time-dependent optimized coupled electron pair approximation (TD-OCEPA0) \cite{pathak2020timedependent}.
This method is much more efficient than time-dependent optimized coupled-cluster with double excitations (TD-OCCD), even though both scale as $O(N^6)$.
The superior efficiency is due to the linear Hermitian structure of the Lagrangian, which helps to avoid a separate solution for the de-excitation amplitudes.
The above argument is also true for the time-dependent
optimized second-order many-body perturbation (TD-OMP2) methods, which scales as $O(N^5)$, reported in Ref. \citenum {pathak2020timemp2}.
{\color{black} We have also reported the implementation and numerical assessments of the second-order approximation to the time-dependent coupled-cluster single double
(TD-CCSD) method, called the TD-CC2 method\cite{pathak2020study}.
The stationary variant of this method (CC2)\cite{christiansen1995second} is one of the widely accepted lower-cost methods.} 
However, the TD-CC2 method suffers from the
lack of gauge invariance and instability in the real-time propagation under the high-intensity fields,
both due to imperfect optimization of orbitals as opposed to the case in fully-optimized TD-OMP2\cite{pathak2020timemp2}.

Thus, the TD-OCCD method and its lower-cost approximations
such as TD-OCEPA0 \cite{pathak2020timedependent} and TD-OMP2 \cite{pathak2020timemp2} will be useful to investigate multielectron dynamics in large chemical systems driven by intense laser fields.
Nevertheless, the inclusion of triple excitation amplitudes is
certainly attractive to achieve decisive accuracy, 
as demonstrated by a quantitative agreement of TD-OCCDT and TD-CASSCF results for strong-field ionization from Ar atom
\cite{sato2018communication}.
Therefore it is desirable to reduce the computational cost of a method including the effect of triple excitations. 
Numerous reports are available in the stationary coupled-cluster literature for reducing the computational scaling retaining a part of the triple excitation amplitudes
\cite{lee1984coupled, urban1985towards, noga1987towards, noga1987full, watts1993coupled}.
{\color{black} The coupled-cluster theory and the many-body perturbation theory are intricately connected.
Therefore, approximation of the coupled-cluster effective Hamiltonian based on many-body perturbation theory is an appealing one. \cite{shavitt:2009}} 

In this article, we report an efficient approximation to the TD-OCCDT method, treating triple excitation amplitudes within the flexibly chosen active space
correct up to the fourth order in the many-body 
perturbation theory. 
We designate this method as TD-OCCDT(4). It is 
closely related to the stationary CCSDT1 method of Bartlett and coworkers \cite{lee1984coupled, urban1985towards}.
We do not include single excitation amplitudes following our previous work\cite{sato2018communication, pathak2020timedependent, pathak2020timemp2, pathak2020study}
{\color{black} but} optimize the orbitals according to TDVP,
which is ideally suited for studying strong-field dynamics involving substantial excitation and ionization.
The TD-OCCDT(4) method inherits gauge invariance from the parent TD-OCCDT method
due to the variational optimization of orbitals.
This method scales as $O(N^7)$, which is an order of magnitude lower than the full TD-OCCDT method, rendering this method very much affordable, while retaining important contributions of the triples.

We apply our newly developed TD-OCCDT(4) scheme to Ne and Ar atoms exposed to a strong near-infrared laser field. 
We compare TD-OCCDT(4) results with those of the TD-CASSCF, TD-OCCDT, TD-OCCD, and uncorrelated TDHF methods to
asses the numerical performance in describing laser-driven multielectron dynamics.
{\color{black} We find that the TD-OCCDT(4) method performs equally as well as the TD-OCCDT and fully correlated TD-CASSCF methods, while reducing the computational cost.}

This paper is organized as follows. We derscribe the TD-OCCDT(4) method in Sec. \ref{sec2}. Section \ref{sec3} presents its numerical application to Ne and Ar. Conclusions are given in Sec.~\ref{sec4}. Hartree atomic units are used unless otherwise stated, and Einstein convention is implied throughout for
summation over orbital indices.
\section{Method}\label{sec2}
\subsection{Background}\label{sec2-1}
The time-dependent electronic Hamiltonian is given by,
\begin{eqnarray}\label{eq:ham1q}
H &=& \sum_{i=1}^{N_e} h(\bm{r}_i,\bm{p}_i, t) + \sum_{i=1}^{N_e-1}\sum_{{\color{black}j=i+1}}^{N_e} \frac{1}{|\bm{r}_i-\bm{r}_j|},
\end{eqnarray}
where $N_e$ denotes the number of electrons, $\bm{r}_i$ and $\bm{p}_i$ the position and canonical momentum, respectively,
of an electron $i$, 
{\color{black}and $h$ the one-electron Hamiltonian operator including the laser-electron interaction.}

In the second quantization representation, it reads
\begin{eqnarray}\label{eq:ham2q}
\hat{H}
&=& h^\mu_\nu \hat{c}^\dagger_\mu\hat{c}_\nu +
 \frac{1}{2}u^{\mu\gamma}_{\nu\lambda} \hat{c}^\dagger_\mu\hat{c}^\dagger_\gamma\hat{c}_\lambda\hat{c}_\nu \\
\label{eq:ham2q-bis}
&=& h^\mu_\nu \hat{E}^\mu_\nu +
 \frac{1}{2}u^{\mu\gamma}_{\nu\lambda} \hat{E}^{\mu\gamma}_{\nu\lambda},
\end{eqnarray}
where $\hat{c}^\dagger_\mu$ ($\hat{c}_\mu$) represents a creation
(annihilation) operator in a complete,
orthonormal set
{\color{black} of $2n_{\rm bas}$ spin-orbitals $\{\psi_\mu(t)\}$, which are, in general, time-dependent, and Eq.~(\ref{eq:ham2q-bis}) defines one-electron ($\hat{E}^\mu_\nu$) and two-electron ($\hat{E}^{\mu\gamma}_{\nu\lambda}$) generators. 
$n_{\rm bas}$ is the number of basis functions used for expanding the spatial part of  $\psi_\mu$}, which, in the present real-space implementation, corresponds to the number of grid points (See Sec.~\ref{sec3}), and 
\begin{eqnarray}
h^\mu_\nu = \int dx_1 \psi^*_\mu(x_1) h(\bm{r}_1,\bm{p}_1) \psi_\nu(x_1),
\end{eqnarray}
\begin{eqnarray}
 u^{\mu\gamma}_{\nu\lambda} = \int\int dx_1dx_2
 \frac{\psi^*_\mu(x_1)\psi^*_\gamma(x_2)\psi_\nu(x_1)\psi_\lambda(x_2)}{|\bm{r}_1-\bm{r}_2|},
\end{eqnarray}
where $x_i=(\bm{r}_i,\sigma_i)$ represents a composite spatial-spin coordinate.

The complete set of $2n_{\rm bas}$ spin-orbitals (labeled with
$\mu,\nu,\gamma,\lambda$) is divided into $n_{\rm occ}$ {\it occupied} ($o,p,q,r,s$) and
$2n_{\rm bas}-n_{\rm occ}$ {\it virtual} spin-orbitals. 
The coupled-cluster (or CI) wavefunction is constructed only with occupied spin-orbitals which are time-dependent in general, and virtual spin-orbitals form the orthogonal complement of the occupied spin-orbital space. 
%
The occupied spin-orbitals
are classified into $n_{\rm core}$ {\it core} spin-orbitals 
which are occupied in the reference $\Phi$ and kept uncorrelated, and
$N=n_{\rm occ}-n_{\rm core}$ {\it active} spin-orbitals ($t,u,v,w$) among which the active electrons are correlated. The active
spin-orbitals are further split into those in the {\it hole} space
($i,j,k,l$) and the {\it particle} space ($a,b,c,d$), which are defined as those occupied and
unoccupied, respectively, in the reference $\Phi$.
The core spin-orbitals can further be split into {\it frozen-core} space ($i^{\prime\prime},j^{\prime\prime}$)
and the {\it dynamical-core} space ($i^\prime,j^\prime$).
The frozen-core orbitals are fixed in time, whereas dynamical core orbitals are propagated in time.\cite{sato2013time}
(See Fig.~1 of Ref.~\citenum{sato2018communication} for a pictorial illustration of the orbital subspacing.)

\subsection{Review of TD-OCC method}\label{sec2-1}
Following our previous work \cite{sato2018communication}, we rely on 
the real action formulation of the time-dependent variational principle with orthonormal
orbitals,
\begin{eqnarray}
S &=& \label{eq:action}
\operatorname{Re}\int_{t_0}^{t_1} Ldt = \frac{1}{2} \int_{t_0}^{t_1}\left( L + L^*\right) dt,
\end{eqnarray}
\begin{eqnarray}
L &=&\label{eq:Lag}
 \langle\Phi|(1+\hat{\Lambda})e^{-\hat{T}}(\hat{H}-i\frac{\partial}{\partial
 t})e^{\hat{T}}|\Phi\rangle,
\end{eqnarray}
where
\begin{eqnarray}\label{eq:amplitude}
\hat{T}&=&\hat{T}_2+\hat{T}_3\cdots=\tau^{ab}_{ij}\hat{E}^{ab}_{ij}+\tau^{abc}_{ijk}\hat{E}^{abc}_{ijk}\cdots, \\
\hat{\Lambda}&=&\hat{\Lambda}_2+\hat{\Lambda}_3\cdots=\lambda_{ab}^{ij}\hat{E}_{ab}^{ij}+\lambda_{abc}^{ijk}\hat{E}_{abc}^{ijk}\cdots,
\end{eqnarray}
where $\tau^{ab\cdots}_{ij\cdots}$ and $\lambda_{ab\cdots}^{ij\cdots}$ are excitation and deexcitation amplitudes, respectively.
The stationary conditions,  $\delta S=0$, 
with respect to
the variation of amplitudes $\delta\tau^{ab\cdots}_{ij\cdots}$,
$\delta\lambda_{ab\cdots}^{ij\cdots}$ and orthonormality-conserving
orbital variations $\delta\psi_\mu$, gives
us the corresponding equations of motions (EOMs).

The coupled-cluster Lagrangian can be written down into two 
equivalent forms, 
\begin{subequations}\label{eqs:Lag_equiv}
\begin{eqnarray}
L&=&
L_0 +
\langle\Phi|(1+\hat{\Lambda})[\{\hat{H}-i\hat{X}\}e^{\hat{T}}]_c|\Phi\rangle
- i\lambda^{ij\cdots}_{ab\cdots}\dot{\tau}^{ab\cdots}_{ij\cdots},
\nonumber \\ \label{eq:Lag_amp}\\
&=&\label{eq:Lag_dens} 
(h^p_q-iX^p_q)\rho^q_p + \frac{1}{2}u^{pr}_{qs} \rho_{pr}^{qs}
- i\lambda^{ij\cdots}_{ab\cdots}\dot{\tau}^{ab\cdots}_{ij\cdots},
\end{eqnarray}
\end{subequations}
where $\hat{X}=X^\mu_\nu\hat{E}^\mu_\nu$, and
$X^\mu_\nu=\langle\psi_\mu|\dot{\psi}_\nu\rangle$ is anti-Hermitian,
$L_0 = \langle\Phi|(\hat{H}-i\hat{X})|\Phi\rangle$ is the reference contribution,
and $[\cdots]_c$ {\color{black} indicates that only the contributions from connected coupled-cluster diagrams 
are retained.}
The one-electron and two-electron reduced density matrices
(RDMs) $\rho^q_p$ and $\rho^{qs}_{pr}$ are defined, respectively, by 
\begin{eqnarray}
 \rho^q_p&=&\label{eq:1rdm}
\langle\Phi|(1+\hat{\Lambda})e^{-\hat{T}}\hat{E}^p_qe^{\hat{T}}|\Phi\rangle, \\
 \rho^{qs}_{pr}&=&\label{eq:2rdm}
\langle\Phi|(1+\hat{\Lambda})e^{-\hat{T}}\hat{E}^{pr}_{qs}e^{\hat{T}}|\Phi\rangle.
\end{eqnarray}
The one-electron and
two-electron RDMs are separated into reference and correlation
contributions, 
\begin{eqnarray}
\rho^q_p &=& (\rho_0)^q_p + \gamma^q_p, \\
\rho^{qs}_{pr}&=& (\rho_0)^{qs}_{pr} + \gamma^{qs}_{pr},
\end{eqnarray}
where the reference contributions
$(\rho_0)^q_p = \delta^q_j\delta^j_p$
and $(\rho_0)^{qs}_{pr}= 
\gamma^q_p \delta^s_j \delta^j_r
+\gamma^s_r \delta^q_j\delta^j_p
-\gamma^q_r \delta^s_j\delta^j_p
-\gamma^s_p \delta^q_j\delta^j_r
+\delta^q_j\delta^j_p\delta^s_k\delta^k_r
-\delta^s_j\delta^j_p\delta^q_k\delta^k_r$
($j,k$ running over core and hole spaces)
are independent of the correlation treatment, and the correlation
contributions are defined as
\begin{subequations} \label{eqs:RDMs_corr}
\begin{eqnarray}\label{eqs:1RDM_corr}
 \gamma^q_p&=&\label{eqs:1RDM_corr}
\langle\Phi|(1+\hat{\Lambda})[\{\hat{E}^p_q\}e^{\hat{T}}]_c|\Phi\rangle, \\
 \gamma^{qs}_{pr}&=&\label{eqs:2RDM_corr}
\langle\Phi|(1+\hat{\Lambda})[\{\hat{E}^{pr}_{qs}\}e^{\hat{T}}]_c|\Phi\rangle,
\end{eqnarray}
\end{subequations}
where the bracket $\{\cdots\}$ implies that the operator
inside is normal ordered relative to the reference.

\subsection{TD-OCCDT(4) method}\label{sec2-2}
The full TD-OCCDT Lagrangian is given by\cite{sato2018communication}
\begin{eqnarray}\label{eq:L_ccd_lin_t}
L_{\rm CCDT}&=&L_0+\langle \Phi|(1+\hat \Lambda_2+\hat \Lambda_3)\nonumber\\
&&[\{\hat H-i\hat X\}(1+\hat T_2+\frac{1}{2!}\hat T_2^2+\hat T_3+\hat T_2 \hat T_3)]_c|\Phi\rangle\nonumber \\
&-&i\lambda_{ab}^{ij}\dot \tau_{ij}^{ab}-i\lambda_{abc}^{ijk}\dot \tau_{ijk}^{abc}, 
\end{eqnarray}
where we truncate after $\hat \Lambda=\hat \Lambda_2+\hat \Lambda_3$ and $\hat T=\hat T_2+\hat T_3$,
expand $e^{\hat{T}_2+\hat{T}_3}$, and retain only the connected diagrams.
For deriving the TD-OCCDT(4) method, we make a further approximation to $L_{\rm CCDT}$ by retaining
those terms contributing up to fourth order in the many-body perturbation theory,
\begin{eqnarray}\label{eq:L_ccdt4}
L_{\rm CCDT}^{(4)}&=&L_0+\langle\Phi|(1+\hat \Lambda_2)[(\bar f+\hat v) e^{\hat T_2}]_c|\Phi\rangle \nonumber \\
&+&\langle\Phi|\hat \Lambda_2 [(\bar f+\hat v)\hat T_3]_c|\Phi\rangle+\langle \Phi|\hat \Lambda_3 (\bar f\hat T_3)_c|\Phi\rangle\nonumber\\
&+&\langle \Phi|\hat \Lambda_3(\hat v\hat T_2)_c|\Phi\rangle-i\lambda_{ab}^{ij}\dot \tau_{ij}^{ab}-i\lambda_{abc}^{ijk}\dot
\tau_{ijk}^{abc},
\end{eqnarray}
where $\bar{f}=\hat{f}-i\hat{X}$, 
$\hat{f} = (h^p_q + v^{pj}_{qj}) \{\hat{E}^p_q\}$,
$\hat{v}=v^{pr}_{qs}\{\hat{E}^{pr}_{qs}\}/4$, 
and $v^{pr}_{qs} = u^{pr}_{qs}-u^{pr}_{sq}$.

Using $L^{(4)}_{\rm CCDT}$ of Eq.~(\ref{eq:L_ccdt4}), 
the TD-OCCDT(4) amplitude equations are derived from the
stationary conditions
$\delta S/\delta \lambda^{ij}_{ab}(t) = 0$, $\delta S/\delta \lambda^{ijk}_{abc}(t) = 0$,
$\delta S/\delta \tau^{ab}_{ij}(t)= 0$, and $\delta S/\delta \tau^{abc}_{ijk}(t) = 0$ as
\begin{eqnarray}
i\dot{\tau}^{ab}_{ij}
&=&\label{eq:td-occd_lint3_t2}
v_{ij}^{ab}-p(ij) \bar{f}_j^k\tau_{ik}^{ab}+p(ab) \bar{f}_c^a \tau_{ij}^{cb} \nonumber \\
&+&\frac{1}{2} v_{cd}^{ab}\tau_{ij}^{cd}
+\frac{1}{2} v_{ij}^{kl} \tau_{kl}^{ab}+p(ij)p(ab)
v_{ic}^{ak} \tau_{kj}^{cb} \nonumber \\ 
&-&\frac{1}{2}p(ij) \tau_{ik}^{ab} \tau_{jl}^{cd} v_{cd}^{kl}
+\frac{1}{2}p(ab) \tau_{ij}^{bc} \tau_{kl}^{ad} v_{cd}^{kl} \nonumber \\
&+&\frac{1}{4} \tau_{kl}^{ab} \tau_{ij}^{cd} v_{cd}^{kl}
+\frac{1}{2}p(ij)p(ab) \tau_{il}^{bc} \tau_{jk}^{ad} v_{cd}^{kl}\nonumber\\
&+&\bar{f}_c^k\tau_{ijk}^{abc}+\frac{1}{2}p(ab)v_{cd}^{ka}\tau_{kij}^{cdb}-\frac{1}{2}p(ij)v_{ci}^{kl}\tau_{klj}^{cab},
\end{eqnarray}
\begin{eqnarray}
i\dot{\tau}^{abc}_{ijk}
&=&\label{eq:td-occd_lint3_t3}
p(k/ij)p(a/bc)v_{dk}^{bc}t_{ij}^{ad}-p(i/jk)p(c/ab)v_{jk}^{lc}t_{il}^{ab}\nonumber \\
&-&p(k/ij)\bar f_k^l\tau_{ijl}^{abc}+p(c/ab)\bar f_d^c\tau_{ijk}^{abd}, 
\end{eqnarray}
\begin{eqnarray}
-i\dot{\lambda}^{ij}_{ab}
&=&\label{eq:td-occd_l2}
v_{ab}^{ij}-p(ij) \bar{f}_k^i
\lambda_{ab}^{kj}+p(ab) \bar{f}_a^c\lambda_{cb}^{ij} \nonumber \\
&+&\frac{1}{2} v_{ab}^{cd}\lambda_{cd}^{ij}
+\frac{1}{2} v_{kl}^{ij}\lambda_{ab}^{kl}+p(ij)p(ab)
v_{kb}^{cj}\lambda_{ac}^{ik} \nonumber \\ 
&-&\frac{1}{2}p(ij) \lambda_{cd}^{ik} \tau^{cd}_{kl} v_{ab}^{jl}
+\frac{1}{2}p(ab) \lambda_{bc}^{kl} \tau^{cd}_{kl} v_{ad}^{ij} \nonumber \\
&+&\frac{1}{4} \lambda_{ab}^{kl} \tau_{kl}^{cd} v_{cd}^{ij}
+\frac{1}{2}p(ij)p(ab) \lambda_{ac}^{jk} \tau_{kl}^{cd} v_{bd}^{il} \nonumber \\
&-&\frac{1}{2}p(ij) \lambda_{ab}^{ik} \tau_{kl}^{cd} v_{cd}^{jl}
+\frac{1}{2}p(ab) \lambda_{bc}^{ij} \tau_{kl}^{cd} v_{ad}^{kl}+\frac{1}{4} \lambda_{cd}^{ij} \tau_{kl}^{cd} v_{ab}^{kl} \nonumber \\
&+&\frac{1}{2}p(ab)v_{bk}^{dc}\lambda_{adc}^{ijk}-\frac{1}{2}p(ij)v_{lk}^{jc}\lambda_{abc}^{ilk},
\end{eqnarray}
\begin{eqnarray}
-i\dot{\lambda}^{ijk}_{abc}
&=&\label{eq:td-occd_l2}
p(k/ij)p(a/bc)v_{bc}^{dk}\lambda_{ad}^{ij}-p(c/ab)p(i/jk)v_{lc}^{jk}\lambda_{ab}^{ij} \nonumber \\
&+&p(c/ab)\bar f_c^d\lambda_{abd}^{ijk}-p(k/ij)\bar f_l^k\lambda_{abc}^{ijl}\nonumber\\
&+&p(i/jk)p(a/bc)\bar f_a^i\lambda_{bc}^{jk},
\end{eqnarray}
where $p(\mu\nu)$ and $p(\mu|\nu\gamma)$ are the permutation operators; ${\color{black}p(\mu\nu)}A_{\mu\nu}=A_{\mu\nu}-A_{\nu\mu}$, and $p(\mu/\nu\gamma)=1-p(\mu\nu)-p(\mu\gamma)$.

Following the discussion in Ref.~\citenum{sato2016time}, the EOM for the orbitals
can be written down in the following form,
\begin{eqnarray}\label{eq:eom_orb}
i|\dot{\psi_p}\rangle &=&
(\hat{1}-\hat{P})
\hat{F}|\psi_p\rangle + i|\psi_q\rangle X^q_p,
\end{eqnarray}
where $\hat{1} = \sum_\mu|\psi_\mu\rangle\langle\psi_\mu|$ is the identity operator within the space spanned by the given basis, 
$\hat{P}=\sum_q|\psi_q\rangle\langle\psi_q|$ is the projector onto the occupied spin-orbital space, and
\begin{eqnarray}
\hat{F}|\psi_p\rangle &=& \label{eq:gfockoperator}
\hat{h} |\psi_p\rangle +  \hat{W}^r_s|\psi_q\rangle
P^{qs}_{or}(D^{-1})_p^o, \\
D^p_q&=&\label{eq:rdmh}\frac{1}{2}(\rho^p_q+\rho^{q*}_p), \\
P^{pr}_{qs}&=&\frac{1}{2}(\rho^{pr}_{qs}+\rho^{qs*}_{pr}),\\
W^r_s(x_1)&=&\label{eq:gfockoperator}
\int dx_2 \frac{\psi^*_r(x_2)\psi_s(x_2)}{|\bm{r}_1-\bm{r}_2|}.
\end{eqnarray}
The matrix element $X^q_p$ describes orbital rotations among various orbital subspaces.
Intra-space orbital rotations ({\color{black}$\{X^{i^\prime}_{j^\prime}\}$},
$\{X^i_j\}$, and $\{X^a_b\}$) are redundant and can be arbitrary
anti-Hermitian matrix elements. On the other hand, inter-space rotations
($X_i^a$ and $X_a^i=-X_i^{a\ast}$) are non-redundant, and determined by solving the
linear equation \cite{sato2018communication}
{\color{black}
\begin{eqnarray}
&i\left(\delta^a_bD^j_i-D^a_b\delta^j_i\right)X^b_j = \label{eq:occ_eom_hp}
B^a_i.&
\end{eqnarray}
The algebraic expressions of RDMs and the matrix $B$ are given in Appendix~\ref{app:density_matrices}.
}
%
{\color{black}
The orbital rotations involving frozen-core orbitals within the electric dipole approximation is given by \cite{sato2016time},
\begin{eqnarray}\label{eq:eom_orb_fa}
iX_\mu^{i^{\prime\prime}}
&=& \left\{
\begin{array}{ll}
 0 & \textrm{(length gauge)}\\
 \pmb{E}(t)\cdot\langle\psi_{i^{\prime\prime}}|\pmb{r}|\psi_\mu\rangle  & \textrm{(velocity gauge)} \\
\end{array}
\right., 
\end{eqnarray}
where $\pmb{E}$ is the external electric field. 
}
\section{Numerical Results and Discussion}\label{sec3}
\begin{figure}[!hb]
\centering
\includegraphics[width=1.1\linewidth]{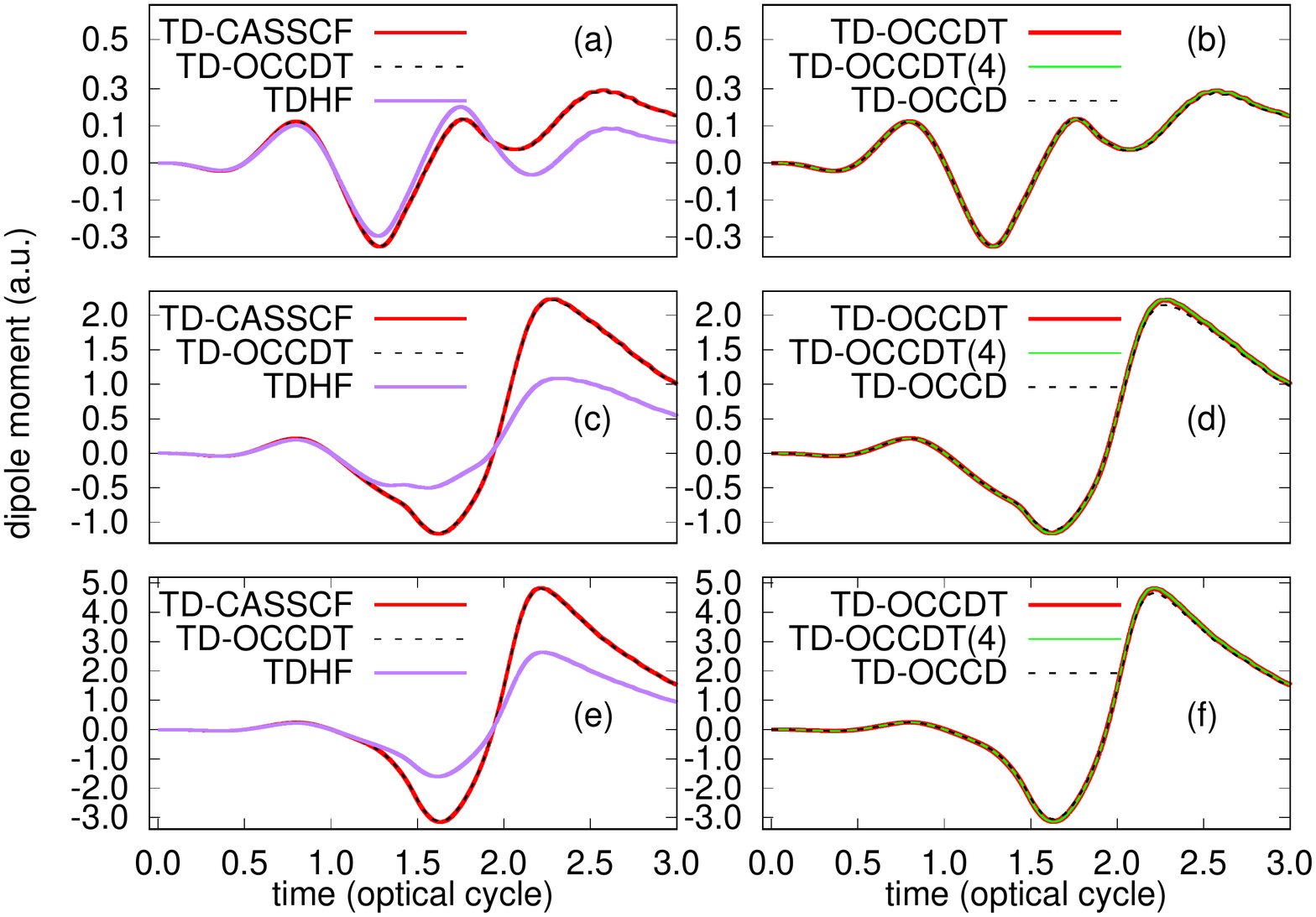}
\caption{\label{fig:nedipole}
{\color{black} Time evolution of dipole moment of Ne irradiated by
a laser pulse with a wavelength of 800 nm and a peak intensity of 5$\times$10$^{14}$ W/cm$^2$ (a, b),
8$\times$10$^{14}$ W/cm$^2$ (c, d), and 1$\times$10$^{15}$ W/cm$^2$ (e, f),
calculated with TDHF, TD-OCCD, TD-OCCDT(4), TD-OCCDT, and TD-CASSCF
methods.}}
\centering
\includegraphics[width=1.1\linewidth]{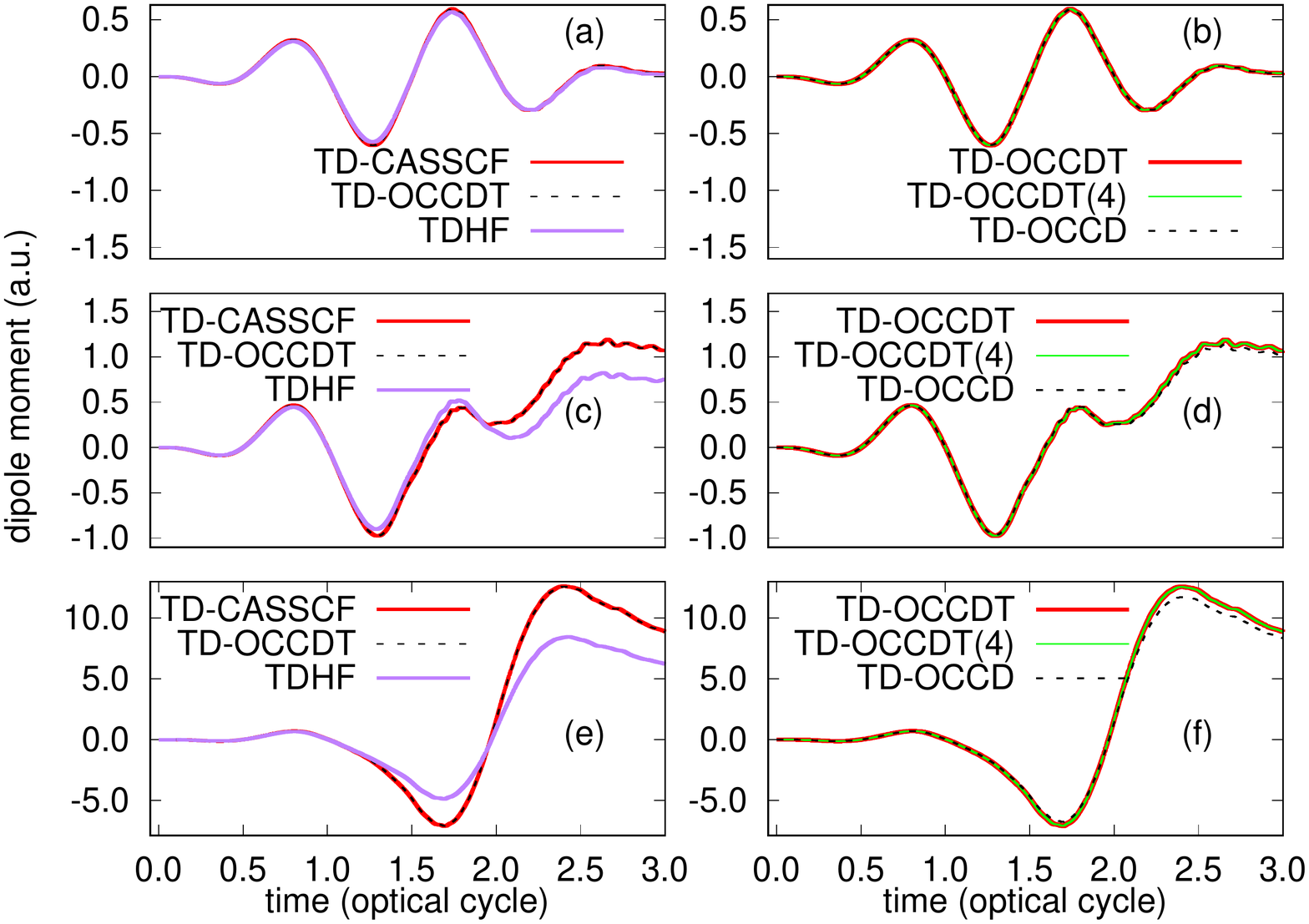}
\caption{\label{fig:ardipole}
{\color{black} Time evolution of dipole moment of Ar irradiated by
a laser pulse with a wavelength of 800 nm and a peak intensity of 1$\times$10$^{14}$ W/cm$^2$ (a, b),
2$\times$10$^{14}$ W/cm$^2$ (c, d) and 4$\times$10$^{14}$ W/cm$^2$ (e, f),
calculated with TDHF, TD-OCCD, TD-OCCDT(4), TD-OCCDT, and TD-CASSCF
methods.}}
\end{figure}
In this section, we apply the TD-OCCDT(4) method to the laser-driven electron dynamics in Ne and Ar. 
The field dependent one-electron Hamiltonian is given by
\begin{eqnarray}\label{eq:h1vg}
h(\bm{r},\bm{p})=\frac{\bm{p}^2}{2} - \frac{Z}{|\bm{r}|} + A(t)p_z,
\end{eqnarray}
within the dipole approximation in the velocity gauge, where
$Z$ is the atomic number, $A(t)=-\int^t E(t^\prime) dt^\prime$ is
the vector potential, with $E(t)$ being the laser electric field
linearly polarized along the $z$ axis. 
Our methods are gauge invariant; both length-gauge and velocity-gauge simulations produce identical results for physical observables upon numerical convergence.
The velocity gauge is known to be advantageous in simulating high-field phenomena \cite{sato2016time,orimo2018implementation, orimo2019tsurff}.

We assume a laser electric field of the following form:
\begin{eqnarray}
E(t)=E_0\,{\text{sin}}(\omega_0t)\,{\text{sin}}^2\left(\pi\frac{t}{3T}\right), 
\end{eqnarray}
for $0 \leq t \leq 3T$, and $E(t)=0$ otherwise, with a central wavelength of $\lambda=2\pi/\omega_0=800$ nm and a period of $T=2\pi/\omega_0 \sim 2.67$ fs. 
We consider three different intensities 5$\times 10^{14}$ W/cm$^2$,
8$\times 10^{14}$ W/cm$^2$, and 1$\times 10^{15}$ W/cm$^2$ for Ne, and 
1$\times 10^{14}$ W/cm$^2$,  2$\times 10^{14}$ W/cm$^2$, and 4$\times 10^{14}$ W/cm$^2$ for Ar.


Our implementation uses a spherical finite-element discrete variable representation (FEDVR) basis \cite{sato2016time, orimo2018implementation} for representing orbital functions,
\begin{eqnarray}
\chi_{klm}(r, \theta, \psi)=\frac{1}{r}f_k(r)Y_{lm}(\theta, \phi) 
\end{eqnarray}
where $Y_{lm}$ and $f_k(r)$ are spherical harmonics and the normalized radial-FEDVR basis function, respectively.
The expansion of the spherical harmonics continued up to the maximum angular momentum $L_{max}$, and the radial FEDVR basis supports the range of radial coordinate $0\leq r \leq R_{max}$,
with cos$^{1/4}$ mask function used as an absorbing boundary for avoiding unphysical reflection from the wall of the simulation box.
We have used $l_{\text{max}}=47$ for Ne and
$l_{\text{max}}=63$ for Ar, and the FEDVR basis supporting the radial
coordinate $0 < r < 260$ using 68 finite elements each containing 21
(for Ne) and 23 (for Ar) DVR functions.
The absorbing boundary switched on at $r = 180$ in all our simulations.
First, the ground state is obtained by the imaginary time relaxation. Starting from the ground state, we perform real-time simulations by switching the laser-field.  
The Fourth-order exponential Runge-Kutta method
\cite{exponential_integrator} is used to propagate the EOMs with 10000
time steps for each optical cycle. 
The simulations are continued for further 3000 time steps after the end of the pulse.
The $1s$ orbital of Ne and $1s2s2p$ orbitals of Ar are kept frozen
at the canonical Hartree-Fock orbitals, and 
for the correlated methods the eight valence electrons are correlated among thirteen active orbitals.

\begin{figure}[!ht]
\centering
\includegraphics[width=1.1\linewidth]{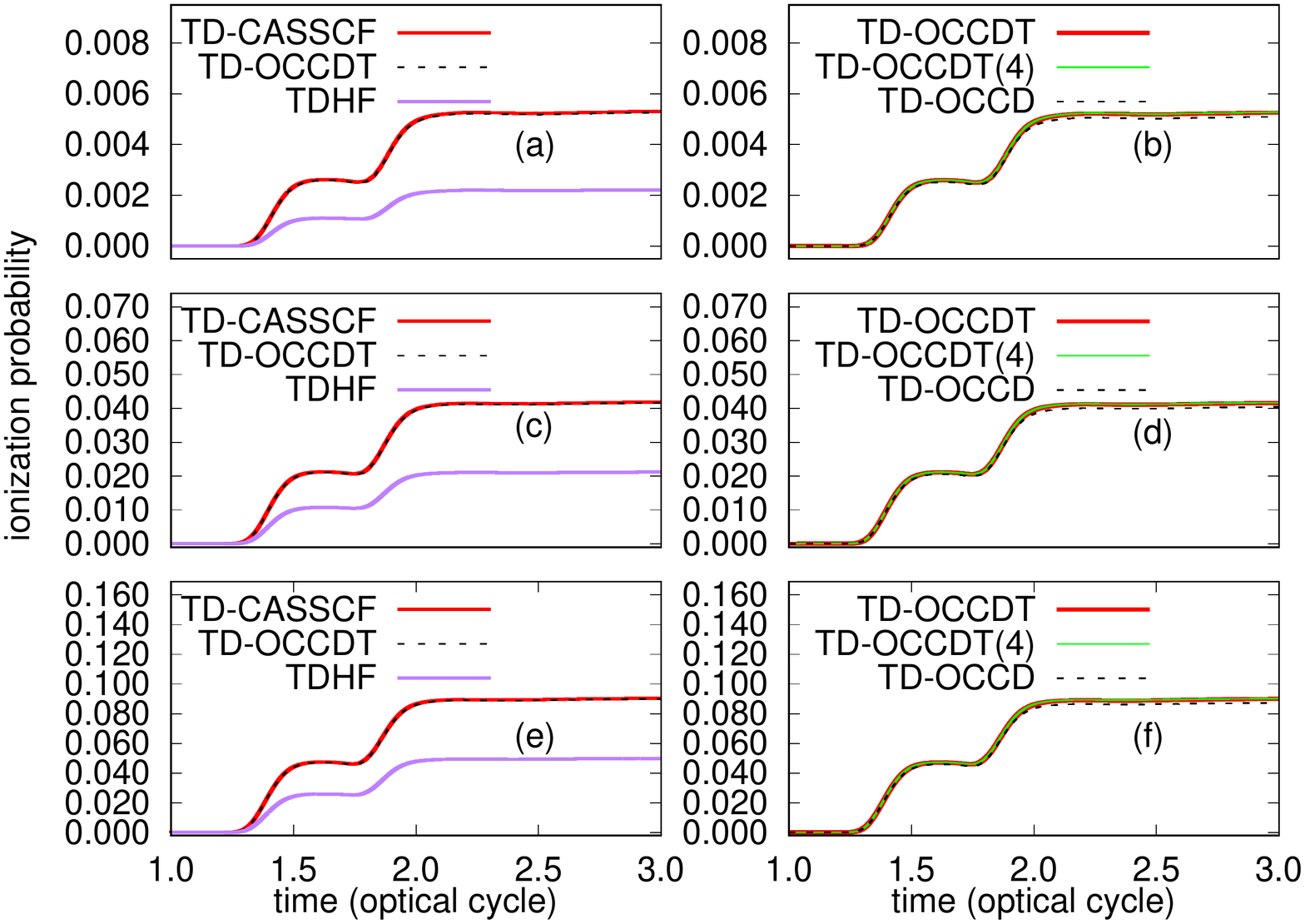}
\caption{\label{fig:nesip}
Time evolution of single ionization probability of Ne irradiated by
a laser pulse with a wavelength of 800 nm and a peak intensity of 5$\times$10$^{14}$ W/cm$^2$ (a, b),
8$\times$10$^{14}$ W/cm$^2$ (c,d) and 1$\times$10$^{15}$ W/cm$^2$ (e, f)
calculated with TDHF, TD-OCCD, TD-OCCDT(4), TD-OCCDT and TD-CASSCF
methods.}
\centering
\includegraphics[width=1.1\linewidth]{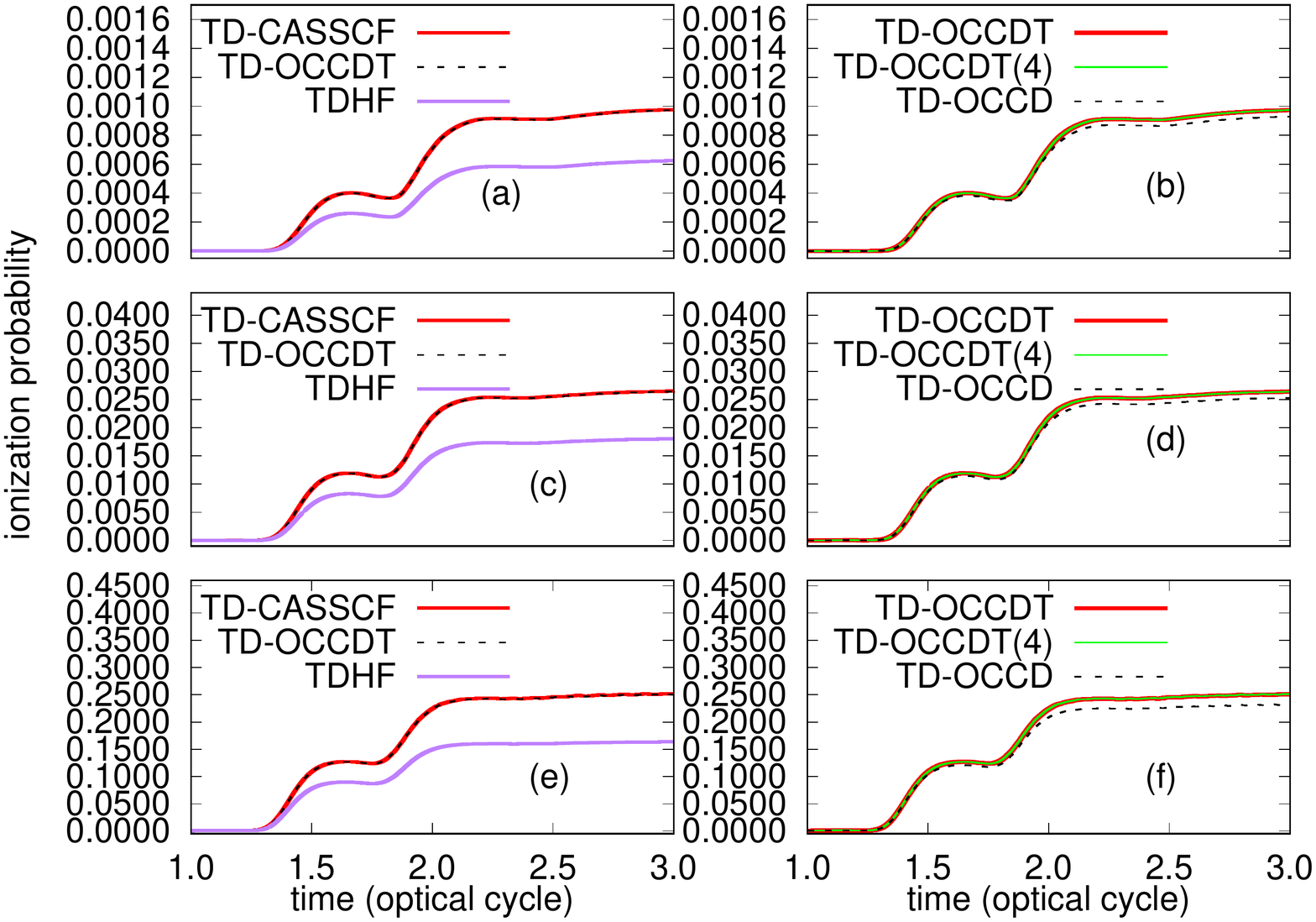}
\caption{\label{fig:arsip}
Time evolution of single ionization probability of Ar irradiated by
a laser pulse with a wavelength of 800 nm and a peak intensity of 1$\times$10$^{14}$ W/cm$^2$ (a, b),
2$\times$10$^{14}$ W/cm$^2$ (c,d) and 4$\times$10$^{14}$ W/cm$^2$ (e, f)
calculated with TDHF, TD-OCCD, TD-OCCDT(4), TD-OCCDT and TD-CASSCF
methods.}
\end{figure}
\begin{figure}[!ht]
\centering
\includegraphics[width=1.1\linewidth]{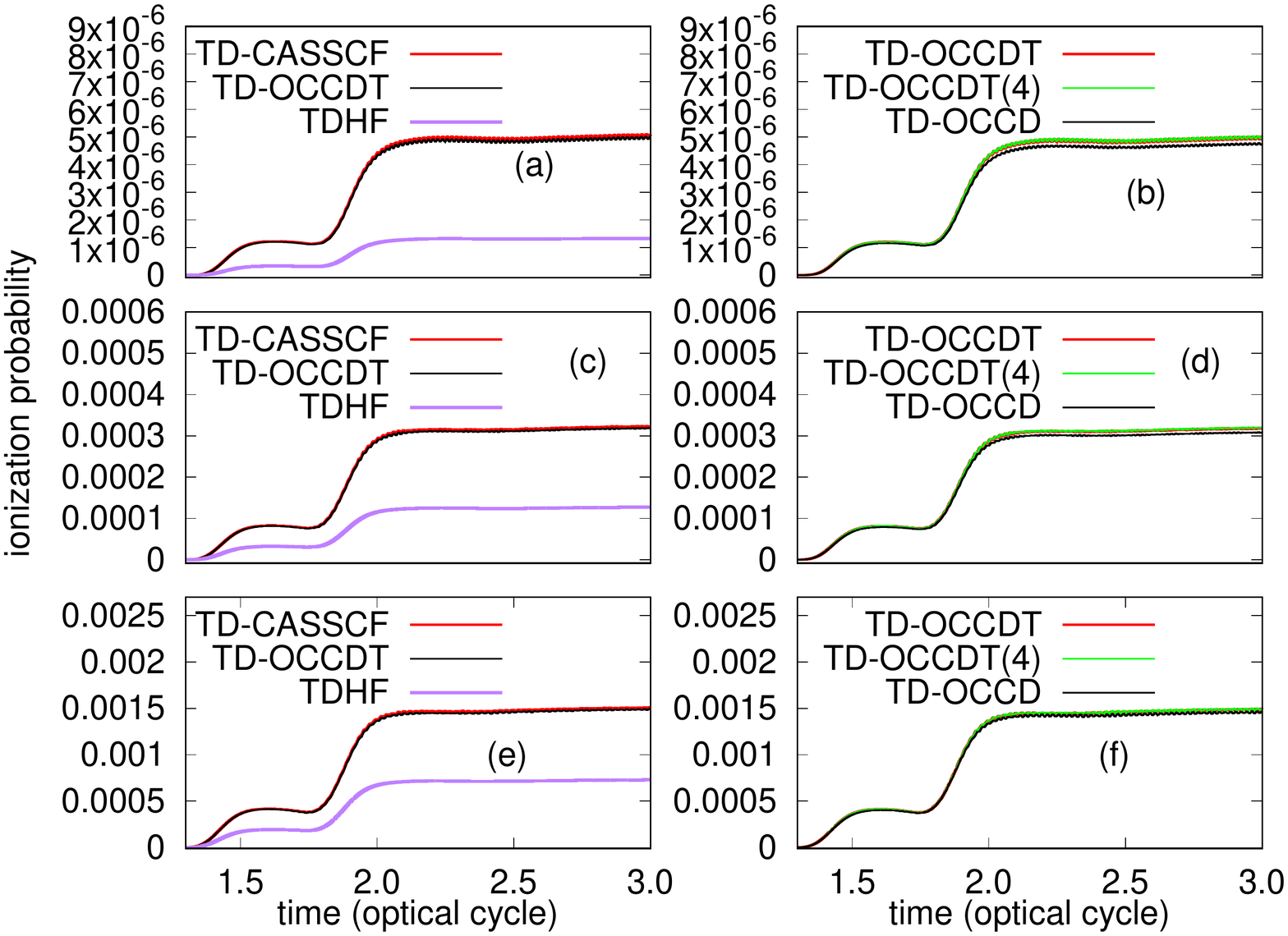}
\caption{\label{fig:nedip}
Time evolution of double ionization probability of Ne irradiated by
a laser pulse with a wavelength of 800 nm and a peak intensity of 5$\times$10$^{14}$ W/cm$^2$ (a, b),
8$\times$10$^{14}$ W/cm$^2$ (c,d) and 1$\times$10$^{15}$ W/cm$^2$ (e, f)
calculated with TDHF, TD-OCCD, TD-OCCDT(4), TD-OCCDT and TD-CASSCF
methods.}
\centering
\includegraphics[width=1.1\linewidth]{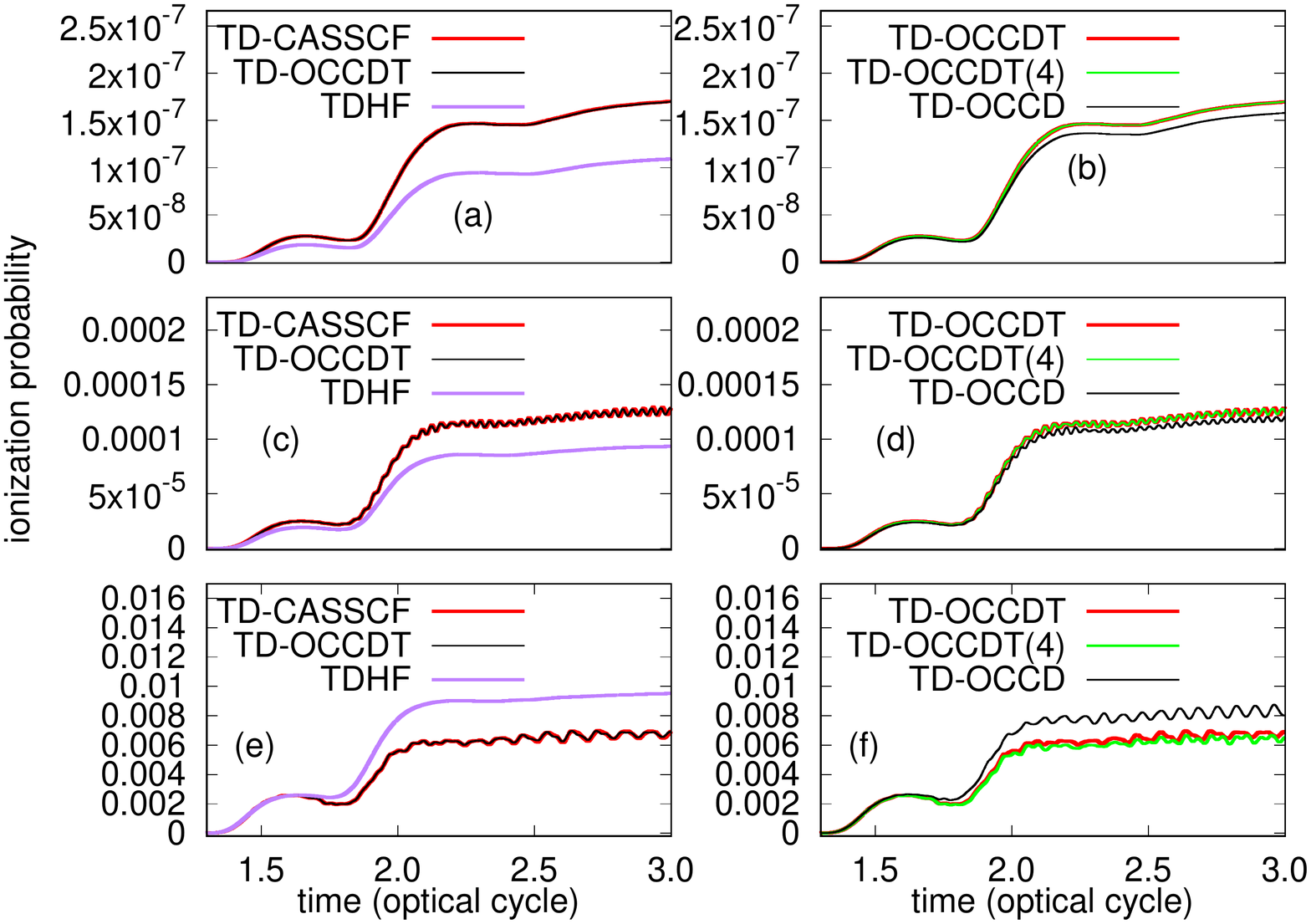}
\caption{\label{fig:ardip}
Time evolution of double ionization probability of Ar irradiated by
a laser pulse with a wavelength of 800 nm and a peak intensity of 1$\times$10$^{14}$ W/cm$^2$ (a, b),
2$\times$10$^{14}$ W/cm$^2$ (c,d) and 4$\times$10$^{14}$ W/cm$^2$ (e, f)
calculated with TDHF, TD-OCCD, TD-OCCDT(4), TD-OCCDT and TD-CASSCF
methods.}
\end{figure}

In Figs.~\ref{fig:nedipole} and \ref{fig:ardipole}, we report
the time evolution of dipole moment, evaluated as a trace $\langle \psi_p|\hat z
|\psi_q\rangle \rho^q_p$ using the 1RDM. 
In Fig. \ref{fig:nedipole} (a)(c)(e), we observe that, whereas the TDHF results deviate substantially from the TD-CASSCF results with larger deviation for higher intensity, the TD-OCCDT produces almost the identical results with those of TD-CASSCF irrespective of the laser intensity.
The valence electrons are driven farther with increasing laser intensity, which renders the dynamical electron correlation (the effect beyond the TDHF description) more relevant.

Figure \ref{fig:nedipole} (b)(d)(f)
compare coupled-cluster results with three different approximation schemes, TD-OCCDT, TD-OCCDT(4), and TD-OCCD.  
We observe that the TD-OCCDT and TD-OCCDT (4) results are indistinguishable on the scale of the figures, while TD-OCCD slightly underestimates the oscillatory behavior. 
The TD-OCCDT(4) method takes account of the dominant part of the electron correlation, successfully approximating the triple excitation amplitudes. 

Figure \ref{fig:ardipole} shows similar results for Ar.
The TDHF, TD-OCCDT, and TD-CASSCF methods give virtually the same results for $10^{14}\,{\rm W/cm}^2$ intensity [Fig.~\ref{fig:ardipole} (a)].
On the other hand, the TDHF result deviates from the others for the higher intensities [Fig.~\ref{fig:ardipole} (c)(e)], indicating that the electron correlation plays an increasingly important role as the electrons are driven farther from the nucleus.
In Fig. \ref{fig:ardipole} (b)(d)(f) we observe, analogous to the Ne case, that the TD-OCCDT(4) result excellently agrees with the TD-OCCDT one, while the TD-OCCD result slightly deviates from them.

Overall, for both Ne (Fig. \ref{fig:nedipole}) and Ar (Fig. \ref{fig:ardipole}),
the TD-OCCDT(4) and TD-OCCDT methods with polynomial scalings deliver practically the same results with the TD-CASSCF method with a factorial scaling, irrespective of the employed intensities.
The TD-OCCDT(4) scales by one order lower than the TD-OCCDT but works equally well with the latter. Thus, the TD-OCCDT(4) method is the most advantageous of the three.


In Figs. \ref{fig:nesip} and \ref{fig:arsip}, we present the temporal evolution of single ionization of Ne and Ar, respectively, evaluated as the probability of finding an electron outside a radius of 20 a.u..
For this purpose we have used RDMs as defined in Refs.~\citenum{fabian2015, fabian2017, sato2018chapter}.
Ionization probabilities are more sensitive to the electron correlation\cite{sato2014structure, hochstuhl2011two}
to assess the capabilities of newly implemented methods in comparison to the established ones.
In panels (a)(c)(e) we compare the TD-CASSCF, TD-OCCDT, and TDHF methods, while we compare different time-dependent optimized coupled-cluster approaches in panels (b)(d)(f).
For both Ne and Ar, the TDHF method leads to substantial underestimation, indicating an important role played by the electron correlation in ionization.
On the other hand, the TD-OCCDT produces nearly identical results with the TD-CASSCF method.
We see from panels (b)(d)(f) that
the TD-OCCDT(4) and TD-OCCDT results are almost
identical to each other, irrespective of the laser intensities, whereas the TD-OCCD results slightly underestimate. 
Especially for Ar [Fig.~\ref{fig:arsip}(b)(d)(f)], the deviation grows with an increasing laser intensity.
This observation again suggests that the consideration of the triples becomes more important for higher intensity due to a greater role of the electron correlation.

Figures \ref{fig:nedip} and \ref{fig:ardip} display the temporal evolution of double ionization of Ne and Ar, respectively, evaluated as the probability of finding two electrons outside a radius of 20 a.u.. 
Double ionization, which may involve recollision, is expected to be even more susceptible to electron correlation.
For the case of Ne (Fig. \ref{fig:nedip}), all the methods except for TDHF predict similar double ionization probability.
However, in the case of Ar (Fig. \ref{fig:ardip}) with more double ionization than for Ne, the TD-OCCD method fails to reproduce the
TD-OCCDT(4) and TD-OCCDT results, which in turn agrees well with the TD-CASSCF.
The deviation is especially large at the highest laser intensity employed [Fig.~\ref{fig:ardip}(f)]. 

\begin{figure}[!ht]
\centering
\includegraphics[width=1\linewidth]{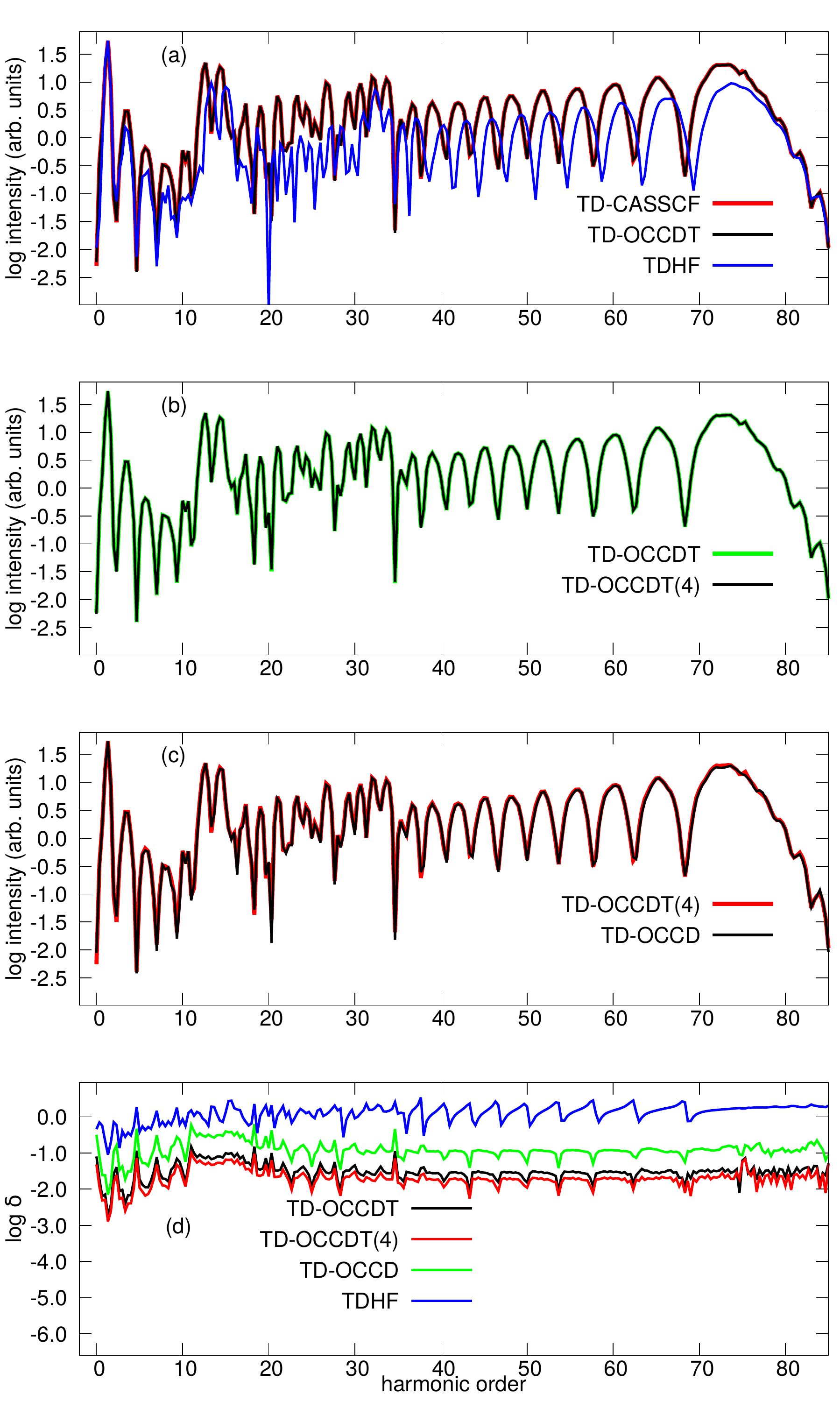}\\
\caption{\label{fig:nehhg5}
The HHG spectra (a, b, c) and the relative deviation (d) of the spectral amplitude from the TD-CASSCF spectrum from Ne irradiated by a laser pulse
with a wavelength of 800 nm and 
a peak intensity of 5$\times$10$^{14}$ W/cm$^2$ with various methods.}
\end{figure}
\begin{figure}[!h]
\centering
\includegraphics[width=1\linewidth]{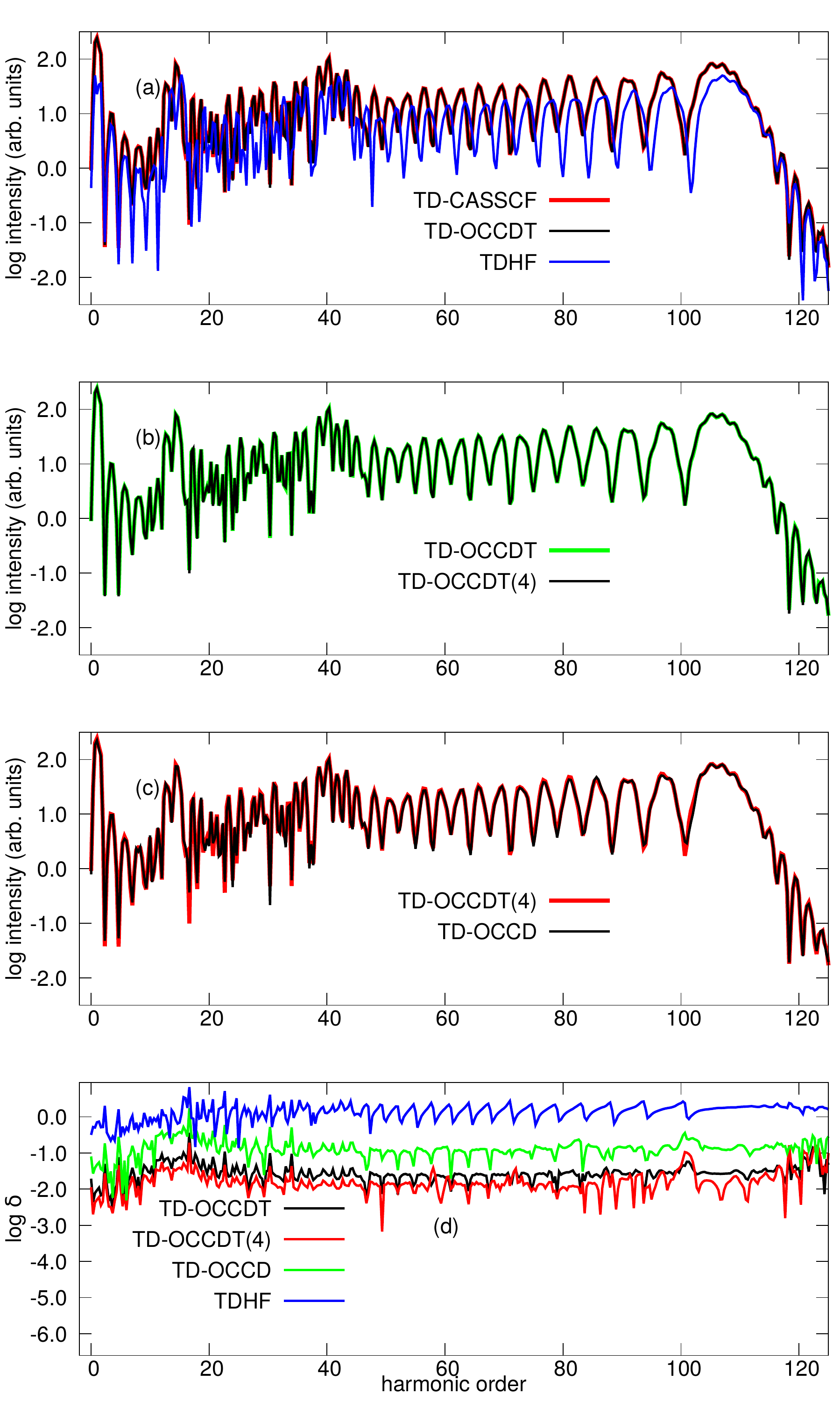}\\
\caption{\label{fig:nehhg8}
The HHG spectra (a, b, c) and the relative deviation (d) of the spectral amplitude from the TD-CASSCF spectrum from Ne irradiated by a laser pulse
with a wavelength of 800 nm and 
a peak intensity of 8$\times$10$^{14}$ W/cm$^2$ with various methods.}
\end{figure}
\begin{figure}[!ht]
\centering
\includegraphics[width=1\linewidth]{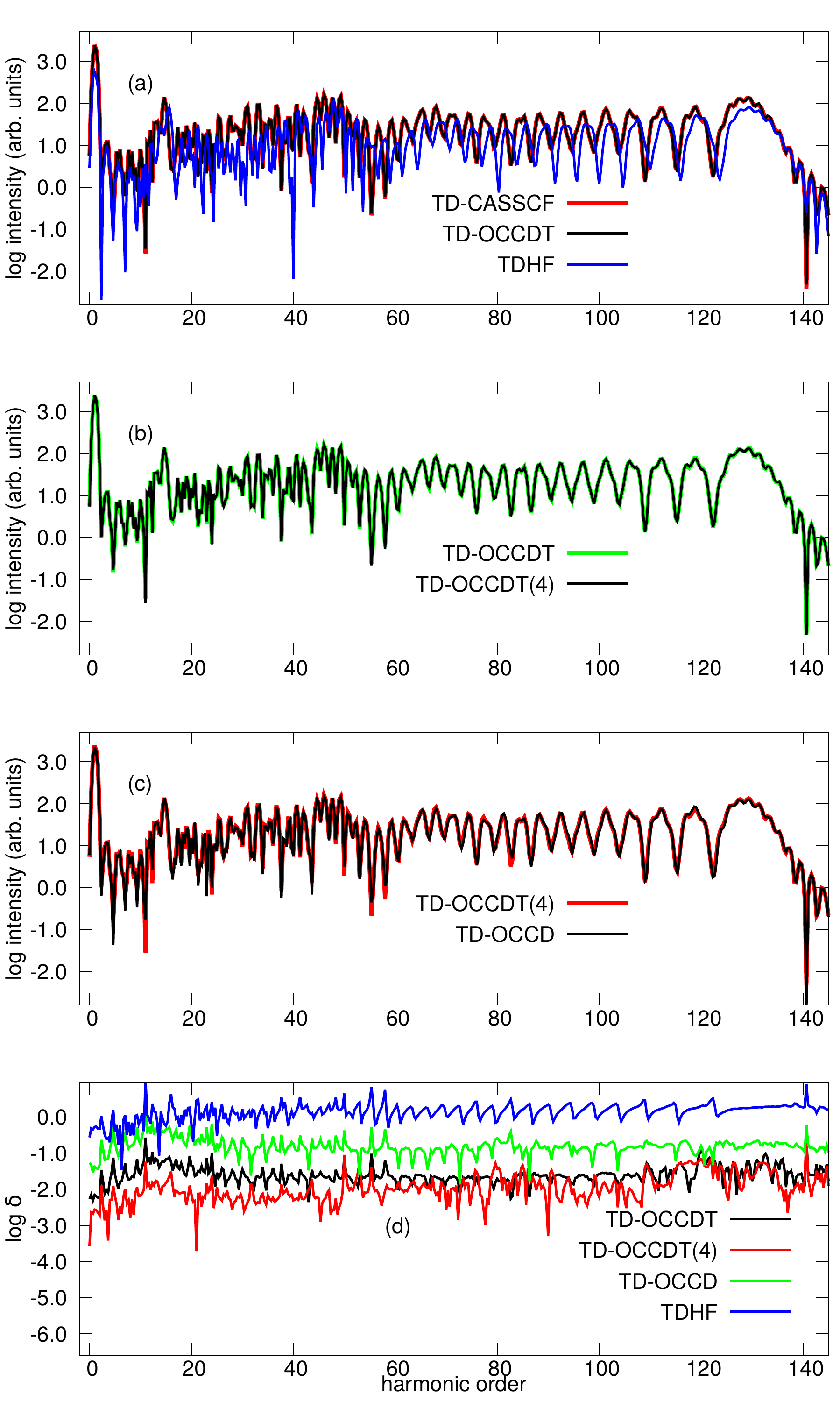}\\
\caption{\label{fig:nehhg10}
The HHG spectra (a, b, c) and the relative deviation (d) of the spectral amplitude from the TD-CASSCF spectrum from Ne irradiated by a laser pulse
with a wavelength of 800 nm and 
a peak intensity of 1$\times$10$^{15}$ W/cm$^2$ with various methods.}
\end{figure}

\begin{figure}[!ht]
\centering
\includegraphics[width=1\linewidth]{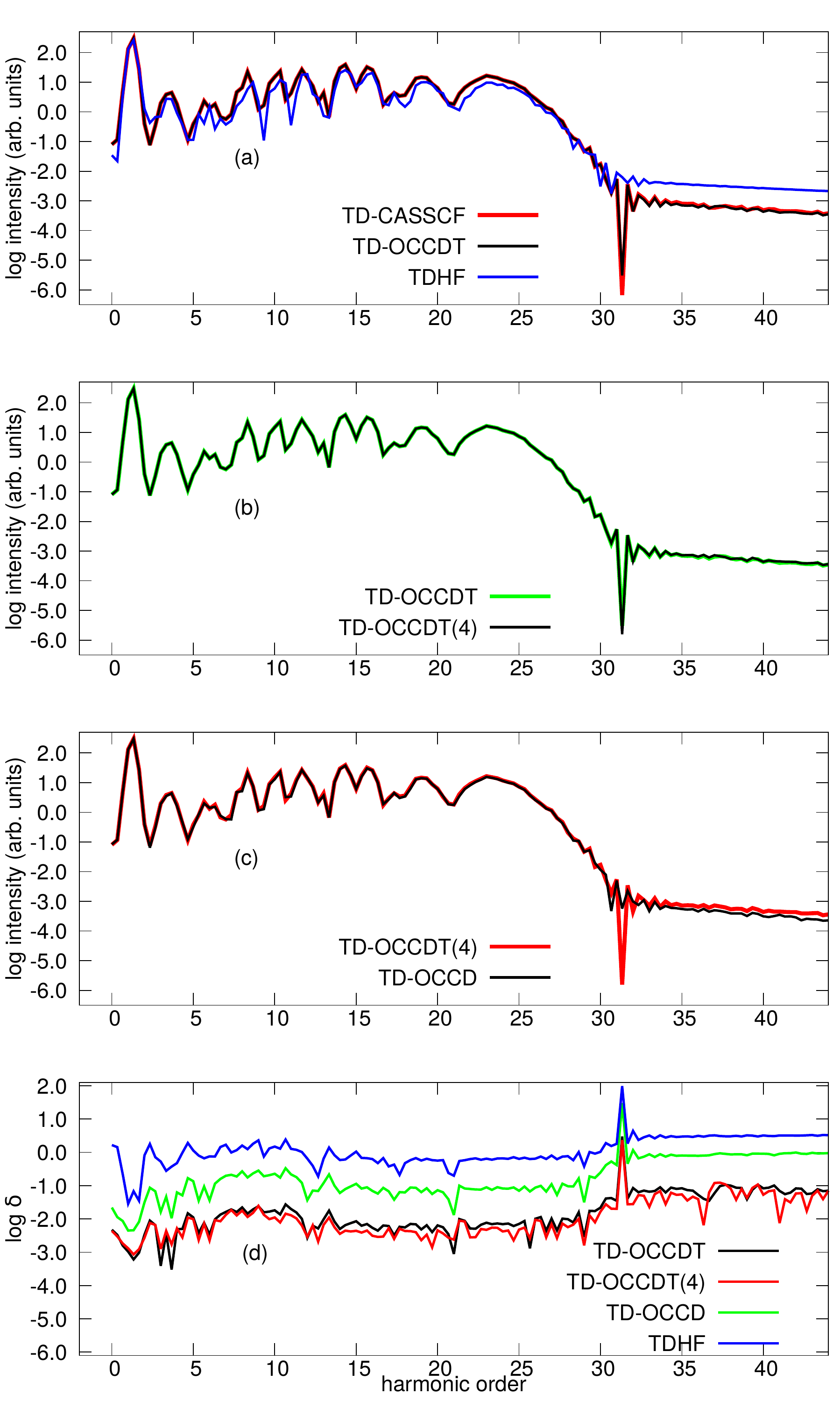}\\
\caption{\label{fig:arhhg1}
The HHG spectra (a, b, c) and the relative deviation (d) of the spectral amplitude from the TD-CASSCF spectrum from Ar irradiated by a laser pulse
with a wavelength of 800 nm and 
a peak intensity of 1$\times$10$^{14}$ W/cm$^2$ with various methods.}
\end{figure}

\begin{figure}[!ht]
\centering
\includegraphics[width=1\linewidth]{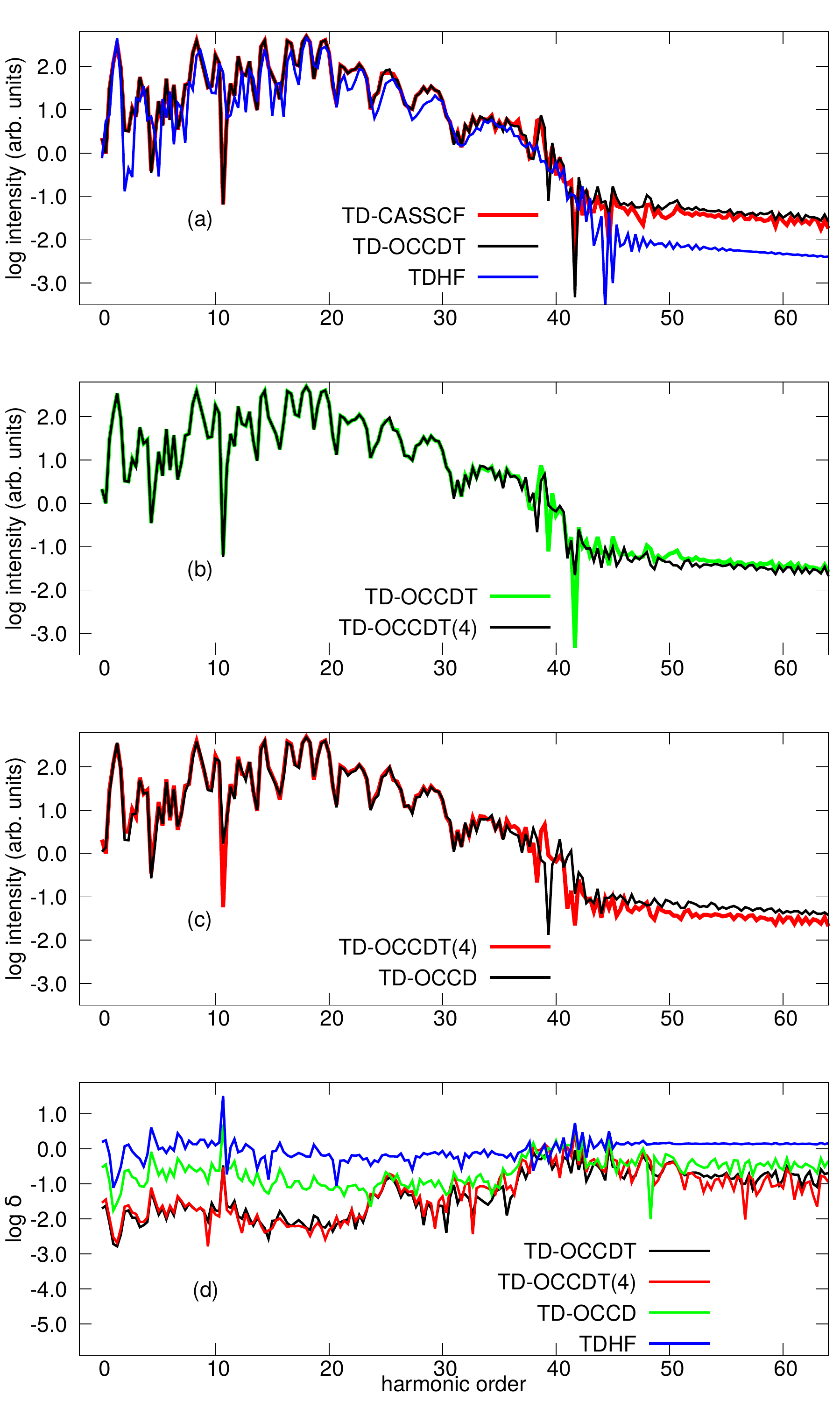}\\
\caption{\label{fig:arhhg2}
The HHG spectra (a, b, c) and the relative deviation (d) of the spectral amplitude from the TD-CASSCF spectrum from Ar irradiated by a laser pulse
with a wavelength of 800 nm and 
a peak intensity of 2$\times$10$^{14}$ W/cm$^2$ with various methods.}
\end{figure}
\begin{figure}[!ht]
\centering
\includegraphics[width=1\linewidth]{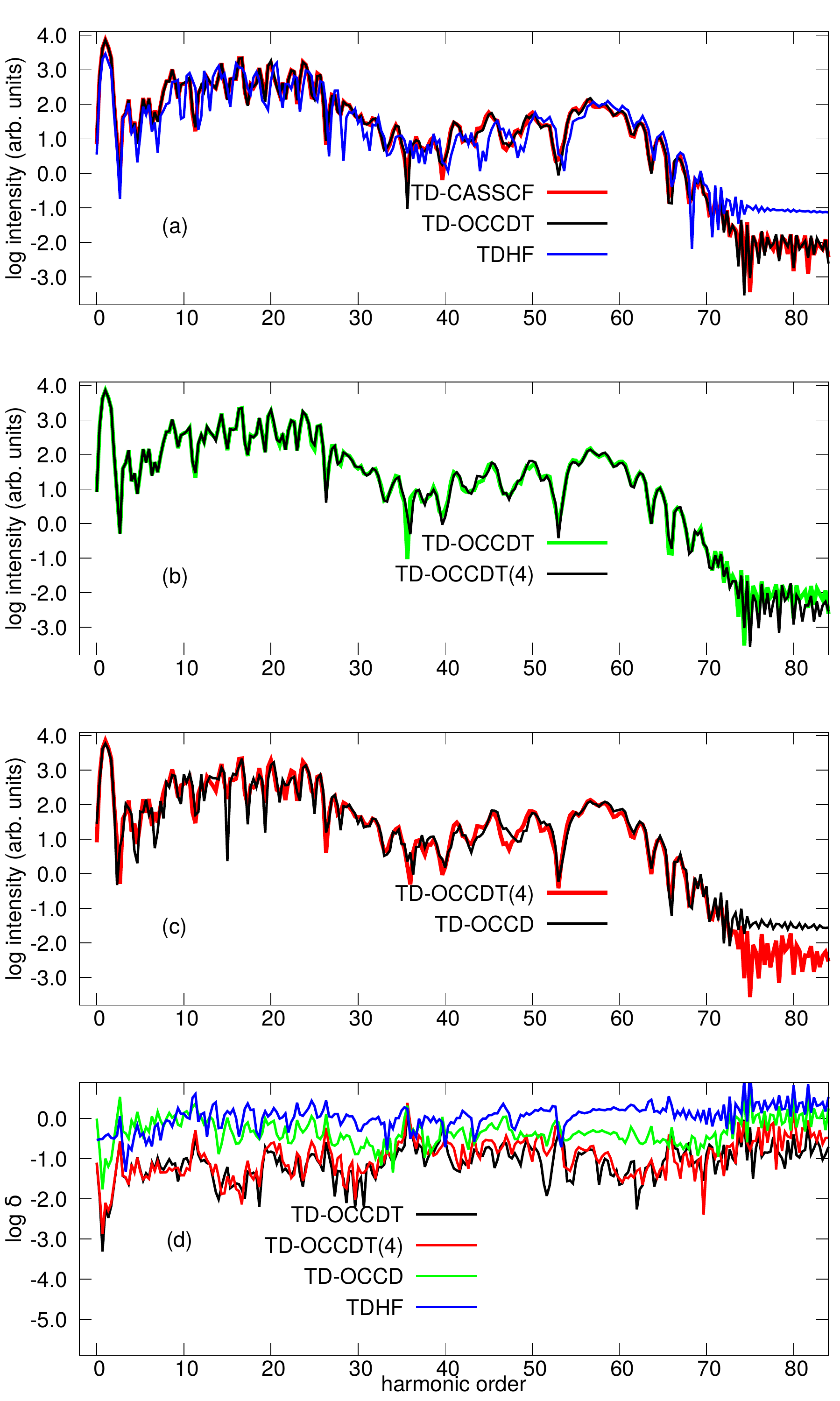}\\
\caption{\label{fig:arhhg4}
The HHG spectra (a, b, c) and the relative deviation (d) of the spectral amplitude from the TD-CASSCF spectrum from Ar irradiated by a laser pulse
with a wavelength of 800 nm and 
a peak intensity of 4$\times$10$^{14}$ W/cm$^2$ with various methods.}
\end{figure}

Let us now turn to high-harmonic generation.
We present the HHG spectra calculated at three different laser intensities for Ne in Figs.~\ref {fig:nehhg5}-\ref {fig:nehhg10} and for Ar in Figs.~\ref {fig:arhhg1}-\ref {fig:arhhg4}.
The HHG spectrum is obtained as the modulus squared $I(\omega) = |a(\omega)|^2$ of the Fourier transform of the expectation
value of the dipole acceleration, which, in turn, is evaluated with a modified Ehrenfest expression \cite{sato2016time}.
We also plot the absolute relative deviation,
\begin{eqnarray}\label{eq:dhhg}
\delta(\omega) = 
\left|
\frac{a(\omega)-a_\textrm{TD-CASSCF}(\omega)}{a_\textrm{TD-CASSCF}(\omega)}
\right|
\end{eqnarray}
of the spectral amplitude $a(\omega)$ from the TD-CASSCF value for each method in the panels (d) of Figs. \ref {fig:nehhg5}-\ref {fig:arhhg4}.  

Again, TDHF underestimates the spectral intensity, presumably due to the underestimation of tunneling ionization [Figs.~\ref{fig:nesip} and \ref{fig:arsip}], i.e.,
the first step in the three-step model\cite{Corkum1993, kulander1993super}. 
The other four methods predict the virtually same spectra on the logarithmic scale of the figures. 
The relative deviation of results for each method from TD-CASSCF ones depends only weakly on the harmonic order and has a general trend of decreasing as TDHF$>$TD-OCCD$>$TD-OCCDT$\gtrsim$TD-OCCDT(4).
This observation clearly demonstrates the usefulness of the TD-OCCDT(4) method.


\begin{table}[h]
\caption{\label{tab:timing_electron} \color{black} Comparison of the total simulation time                                                             
time$^\text{a}$ (in min) spent for TD-CASSCF, TD-OCCDT, TDCCDT(4), and TD-OCCD methods}
\begin{ruledtabular}
\begin{center}
\begin{tabular}{rrr}
Method & \multicolumn{1}{c}{Time (min)} & cost reduction (\%)\\
\hline
TD-CASSCF &\, 7638 &\,\dots \\
TD-OCCDT &\, 5734 &\, 25\\
TD-OCCDT(4) &\, 4600 &\, 40\\ 
TD-OCCD&\,3904&\, 49\\
\end{tabular}
\end{center}
\end{ruledtabular}
\footnotetext[1]                                                                                                                     
{\color{black} Time spent for the simulation of Ar atom for                                                                      
33000 time steps ($0 \leq t \leq 3.3T$) of a real-time simulation                                                                                           
 ($I_0=4\times 10^{14}$ W/cm$^{2}$ and $\lambda=800$ nm.),
 using an Intel(R) Xeon(R)  Gold 6230 CPU with 10 processors having a clock speed of 2.10GHz.}                                
\end{table}

Finally, we compare the computational cost of the different methods considered in this article.
All our simulations have been run on an Intel(R) Xeon(R) Gold 6230 central processing unit (CPU) with 10 processors with a clock speed of 2.10 GHz.
Table \ref{tab:timing_electron} reports the total computation time spent for the simulation corresponding to Fig. \ref{fig:arhhg4}.
This table also lists the relative cost reduction with respect to the TD-CASSCF method. 
The cost reduction for the TD-OCCDT(4) method (40\%) is larger than for the TD-OCCDT method (25\%).
As we have repeatedly seen above, the inclusion of triples is certainly important for highly accurate simulations of high-field phenomena, and the TD-OCCDT(4) method takes account of the essential part of the triples at an affordable reduced computational cost.

\section{Concluding Remarks}\label{sec4}
We have developed the TD-OCCDT(4) method as a cost-effective and efficient approximation to the TD-OCCDT method. 
This method considers triple excitation
amplitudes correct up to the fourth order in the many-body perturbation theory. 
Its computational cost scales as $O(N^7)$.
As numerical assessment, we have applied this method to Ne and Ar atoms irradiated by an intense near-infrared laser pulse and compared the calculated time-dependent dipole moment, single and double ionization probability, and HHG spectra with those obtained with other methods.
We have found that the TD-OCCDT(4) method takes account the major part of triple excitation amplitudes, which are important for obtaining consistently accurate results over a wide range of problems; the TD-OCCDT(4) method delivers the results virtually indistinguishable from those of its parent, full TD-OCCDT method. 
Furthermore, the TD-OCCDT(4) method achieves a 40\% computational cost reduction in comparison to the TD-CASSCF method for the case of Ar calculation with eight active electrons with thirteen active orbitals.
The present TD-OCCDT(4) method will extend the applicability of highly accurate {\it ab initio} simulations of electron dynamics to larger chemical systems.

\section*{acknowledgments}
{\color{black}
This research was supported in part by a Grant-in-Aid for
Scientific Research (Grants No. JP18H03891 and No. JP19H00869) from the Ministry of Education, Culture,
Sports, Science and Technology (MEXT) of Japan. 
This research was also partially supported by JST COI (Grant No.~JPMJCE1313), JST CREST (Grant No.~JPMJCR15N1),
and by MEXT Quantum Leap Flagship Program (MEXT Q-LEAP) Grant Number JPMXS0118067246.}

\section*{DATA AVAILABLITY}
The data that support the findings of this study are available from the corresponding author
upon reasonable request.
\appendix
{\color{black}
\section{Algebraic details of the reduced density matrices and the matrix $B$ for TD-OCCDT(4) method}\label{app:density_matrices}}
The algebraic expressions of non-vanishing correlation contributions to 1RDMs and 2RDMs are given by
\begin{subequations}
\begin{eqnarray}
&&\gamma^{ j }_{ i }=\label{eqs:td-occd_den1}
- \frac{ 1 }{ 2 }
\lambda^{ k j }_{ c d } 
\tau^{ c d }_{ k i }-\frac{1}{12}\lambda_{abc}^{klj}\tau_{kli}^{abc},\\
&&\gamma^{ b }_{ a }=
\frac{ 1 }{ 2 }
\lambda^{ k l }_{ c a } 
\tau^{ c b }_{ k l }+\frac{1}{12}\lambda_{cda}^{ijk}\tau_{ijk}^{cdb},\\
&&\gamma^{a}_{i}=
\frac{1}{4}\lambda_{bc}^{jk}\tau_{jki}^{bca},\\
&&\gamma^{ c d }_{ a b }=
\frac{ 1 }{ 2 }
\lambda^{ k l }_{ a b } 
\tau^{ c d }_{ k l },
\gamma^{ k l }_{ i j }=
\frac{ 1 }{ 2 }
\lambda^{ k l }_{ c d } 
\tau^{ c d }_{ i j }, \\
&&\gamma_{ci}^{ab}=\frac{1}{2}\lambda_{dc}^{jk}\tau_{jki}^{dab}, \gamma_{jk}^{ia}=-\frac{1}{2}\lambda_{bc}^{li}\tau_{ljk}^{bca},\\
&&\gamma^{ i a }_{ b j }=
\lambda^{ k i }_{ c b }
\tau^{ c a }_{ k j }, \gamma_{bc}^{ia}=\frac{1}{2}\lambda_{dbc}^{jki}\tau_{jk}^{da},\\
&&\gamma^{ i j }_{ a b }=
\lambda^{ i j }_{ a b }, \gamma_{bj}^{ik}-\frac{1}{2}\lambda_{cab}^{lik}\tau_{lj}^{ca},\\
&&\gamma^{ a b }_{ i j }=\label{eq:td-occd_den2}
\tau^{ a b }_{ i j }
+\frac{ 1 }{ 2 }
p(ij)p(ab)
\lambda^{ k l }_{ c d }
\tau^{ c a }_{ k i }
\tau^{ b d }_{ j l } \nonumber \\
&&-\frac{ 1 }{ 2 }
p( i j )
\lambda^{ k l }_{ c d }
\tau^{ c d }_{ k i }
\tau^{ a b }_{ l j }
-\frac{ 1 }{ 2 }
p( a b )
\lambda^{ k l }_{ c d }
\tau^{ c a }_{ k l }
\tau^{ d b }_{ i j } \nonumber \\
&&+
\frac{ 1 }{ 4 }
\lambda^{ k l }_{ c d }
\tau^{ c d }_{ i j }
\tau^{ a b }_{ k l }.
\end{eqnarray}
\end{subequations}
{\color{black}
The matrix $B$ is given by
\begin{eqnarray}
B^a_i &=& \label{eq:bvec}
F^a_pD^p_i - D^a_pF^{i*}_p -
\frac{i}{8} \dot \tau_{ijk}^{abc}\lambda_{bc}^{jk} - \frac{i}{8} \tau_{ijk}^{abc}\dot\lambda_{bc}^{jk}, 
\end{eqnarray}
where $F^p_q = \langle\psi_p|\hat{F}|\psi_q\rangle$, with the operator $\hat{F}$ defined by Eq.~(\ref{eq:gfockoperator}).}

%

\end{document}